\RequirePackage{ifpdf}
\ifpdf % We are running pdfTeX in pdf mode
\documentclass[pdftex]{sigma}
\else
\documentclass{sigma}
\fi

\begin{document}

\allowdisplaybreaks

\renewcommand{\thefootnote}{$\star$}

\renewcommand{\PaperNumber}{089}

\FirstPageHeading

\ShortArticleName{Comment on ``Non-Hermitian Quantum Mechanics with Minimal Length Uncertainty''}

\ArticleName{Comment on ``Non-Hermitian Quantum Mechanics\\ with Minimal Length Uncertainty''\footnote{This paper is a
contribution to the Proceedings of the 5-th Microconference
``Analytic and Algebraic Me\-thods~V''. The full collection is
available at
\href{http://www.emis.de/journals/SIGMA/Prague2009.html}{http://www.emis.de/journals/SIGMA/Prague2009.html}}}

\Author{Bijan BAGCHI~$^\dag$ and Andreas FRING~$^\ddag$}

\AuthorNameForHeading{B.~Bagchi and A.~Fring}

\Address{$^\dag$~Department of Applied Mathematics, University of Calcutta,\\
 \hphantom{$^\dag$}~92 Acharya Prafulla Chandra Road, Kolkata 700 009, India}
\EmailD{\href{mailto:BBagchi123@rediffmail.com}{BBagchi123@rediffmail.com}}

\Address{$^\ddag$~Centre for Mathematical Science, City University London,\\
 \hphantom{$^\ddag$}~Northampton Square, London EC1V 0HB, UK}
\EmailD{\href{mailto:A.Fring@city.ac.uk}{A.Fring@city.ac.uk}}

\ArticleDates{Received August 18, 2009;  Published online September 17, 2009}

\Abstract{We demonstrate that the recent paper by Jana and Roy entitled
``Non-Hermitian quantum mechanics with minimal length uncertainty'' [\href{http://dx.doi.org/10.3842/SIGMA.2009.083}{{\it SIGMA} \textbf{5} (2009), 083, 7~pages}, \href{http://arxiv.org/abs/0908.1755}{arXiv:0908.1755}] contains various misconceptions. We compare with an analysis on the same topic carried out previously in our manuscript [\href{http://arxiv.org/abs/0907.5354}{arXiv:0907.5354}]. In particular, we show that the metric operators computed for the deformed non-Hermitian Swanson models dif\/fers in both cases and is inconsistent in the former.}


\Classification{81Q10; 46C15; 81Q12}

\noindent
It is known for some time that the deformations of the standard canonical
commutation relations between the position operator $P$ and the momentum
operator $X$ will inevitably lead to a minimal length, that is a bound
beyond which the localization of space-time events are no longer possible.
In a recent manuscript \cite{BBAF} we investigated various limits of the $q$-deformationed relations
\begin{gather*}
\left[ X,P\right] =i\hbar q^{f(N)}(\alpha \delta +\beta \gamma )+\frac{%
i\hbar (q^{2}-1)}{\alpha \delta +\beta \gamma }\left( \delta \gamma
X^{2}+\alpha \beta ~P^{2}+i\alpha \delta XP-i\beta \gamma PX\right) ,
%\label{gxp}
\end{gather*}%
in conjunction with the constraint $4\alpha \gamma =(q^{2}+1)$, with $\alpha
,\beta ,\gamma ,\delta \in \mathbb{R}$ and $f$ being an arbitrary function
of the number operator $N$. One may consider various types of Hamiltonian
systems, either Hermitian or non-Hermitian, and replace the original
standard canonical variables $(x_{0},p_{0})$, obeying $\left[ x_{0},p_{0}%
\right] =i\hbar $, by $(X,P)$. It is crucial to note that even when the
undeformed Hamiltonian is Hermitian $H(x_{0},p_{0})=H^{\dagger }(x_{0},p_{0})
$ the deformed Hamiltonian is inevitably non-Hermitian $H(X,P)\neq
H^{\dagger }(X,P)$ as a consequence of the fact that $X$ and/or $P$ are no
longer Hermitian. Of course one may also deform Hamiltonians, which are
already non-Hermitian when undeformed $H(x_{0},p_{0})\neq H^{\dagger
}(x_{0},p_{0})$. In both cases a proper quantum mechanical description
requires the re-def\/inition of the metric to compensate for the introduction
of non-Hermitian variables and in the latter an additional change due to the
fact that the Hamiltonian was non-Hermitian in the f\/irst place.

In a certain limit, as specif\/ied in \cite{BBAF}, $X$ and $P$ allow for a
well-known representation of the form $X=(1+\tau p_{0}^{2})x_{0}$ and$%
~P=p_{0}$, which in momentum space, i.e.\ $x_{0}=i\hbar \partial _{p_{0}}$,
corresponds to the one used by Jana and Roy \cite{JanaRoy}, up to an
irrelevant additional term $i\hbar \tilde{\gamma}P$. (Whenever constants
with the same name but dif\/ferent meanings occur in \cite{JanaRoy} and \cite%
{BBAF} we dress the former with a tilde.) The additional term can simply be
gauged away and has no physical signif\/icance. Jana and Roy have studied the
non-Hermitian displaced harmonic oscillator and the Swanson model. As we
have previously also investigated the latter in \cite{BBAF}, we shall
comment on the dif\/ferences. The conventions in \cite{JanaRoy} are
\begin{gather*}
H_{\text{JR}}(a,a^{\dagger })=\omega a^{\dagger }a+\lambda a^{2}+\tilde{%
\delta}(a^{\dagger })^{2}+\frac{\omega }{2}  %\label{JR}
\end{gather*}
with $\lambda \neq \tilde{\delta}\in \mathbb{R}$ and $a=(P-i\omega X)/\sqrt{%
2m\hbar \omega }$, $a^{\dagger }=(P+i\omega X)/\sqrt{2m\hbar \omega }$,
whereas in \cite{BBAF} we used%
\begin{gather*}
H_{\text{BF}}(X,P)=\frac{P^{2}}{2m}+\frac{m\omega ^{2}}{2}X^{2}+i\mu \{X,P\}
%\label{BF}
\end{gather*}%
with $\mu \in \mathbb{R}$ as a starting point. Setting $\hbar =m=1$ it is
easy to see that the models coincide when $\lambda =-\tilde{\delta}$ and $%
\mu =\tilde{\delta}-\lambda $. The Hamiltonians exhibit a $``$%
twofold\textquotedblright\ non-Hermiticity, one resulting from the fact that
when $\lambda \neq \tilde{\delta}$ even the undeformed Hamiltonian is
non-Hermitian and the other resulting from the replacement of the Hermitian
variables $(x_{0},p_{0})$ by $(X,P)$. The factor of the metric operator to
compensate for the non-Hermiticity of $X$ coincides in both cases, but the
factor which is required due to the non-Hermitian nature of the undeformed
case dif\/fers in both cases%
\begin{gather*}
\rho _{\text{BF}}=e^{2\mu P^{2}}\qquad \text{and\qquad }\rho _{\text{JR}%
}=(1+\tau P^{2})^{\frac{\mu }{\omega ^{2}\tau }}.
\end{gather*}
We have made the above identif\/ications such that $H_{\text{JR}}(a,a^{\dagger
})=H_{\text{BF}}(X,P)$ and replaced the deformation parameter $\beta $ used
in \cite{JanaRoy} by $\tau $ employed in \cite{BBAF}. It is well known that
when given only a non-Hermitian Hamiltonian, the metric operator can not be
uniquely determined. However, as argued in \cite{BBAF} with the
specif\/ication of the observable $X$, which coincides in \cite{JanaRoy} and
\cite{BBAF}, the outcome is unique and we can therefore directly compare $%
\rho _{\text{BF}}$ and $\rho _{\text{JR}}$. The limit $\tau \rightarrow 0$
reduces the deformed Hamiltonian $H_{\text{JR}}=H_{\text{BF}}$ to the
standard Swanson Hamiltonian, such that $\rho _{\text{JR}}$ and $\rho _{%
\text{BF}}$ should acquire the form of a previously constructed metric
operator. This is indeed the case for $\rho _{\text{BF}}$, but not for $\rho
_{{\rm JR}}$. In fact it is unclear how to carry out this limit for $\rho _{%
\text{JR}}$ and we therefore conclude that the metric $\rho _{\text{JR}}$ is
incorrect.

\pdfbookmark[1]{References}{ref}
\LastPageEnding

\end{document}